\documentclass{ws-procs9x6}

\begin{document}

\title{Negative Energy Densities in Quantum Field Theory}

\author{L.H. Ford}

\address{Institute of Cosmology  \\
Department of Physics and Astronomy\\ 
         Tufts University, Medford, MA 02155 USA
$^*$E-mail: ford@cosmos.phy.tufts.edu}

\begin{abstract}
Quantum field theory allows for the suppression of vacuum fluctuations,
leading to sub-vacuum phenomena. One of these is the appearance of
local negative energy density. Selected aspects of negative energy
will be reviewed, including the quantum inequalities which limit
its magnitude and duration. However, these
inequalities allow the possibility that negative energy and related
effects might be observable. Some recent proposals for experiments
to search for sub-vacuum phenomena will be discussed. 
Fluctuations of the energy density around its mean value 
will also be considered, and some recent results on a probability
distribution for the energy density in two dimensional
spacetime are summarized.
\end{abstract}

\section{Introduction}

Although the local energy density for the electromagnetic and
other known fields is positive at the classical level, this need
not be the case at the quantum level. When defining the expectation
value of the stress tensor in quantum field theory, a subtraction
is needed, after which a negative value can result. In the case
of empty Minkowski spacetime, we take the vacuum state as the zero
of energy density. Negative values for the mean energy density are
examples of sub-vacuum phenomena, where the effects of vacuum
fluctuations have effectively been suppressed. We will review some
examples of this effect in the Casimir effect and in non-classical
quantum states, and also review the quantum inequalities which limit
sub-vacuum phenomena. In Sect.~\ref{sec:exp}, some proposed laboratory
experiments to detect sub-vacuum effects will be discussed. Finally
we turn to the topic of stress tensor fluctuations, whereby negative
energy can arise by a quantum fluctuation even when the expectation
value of the energy is positive or zero. Some recent results on
the probability distribution for such fluctuations will be discussed.

\section{Negative Energy Density in the Casimir Effect}

Simply from
Casimir's result for the force per unit area~\cite{Casimir} between
a pair of parallel perfectly reflecting plates, 
one can construct the entire
stress tensor, using conservation, tracelessness and symmetry 
arguments~\cite{Brown,DeWitt}. The result, in units where
 $\hbar = c =1$, is
\begin{equation}
T_{\mu\nu} =
\begin{pmatrix} T_{00} & 0 & 0 & 0 \\
                 0 & T_{xx} & 0 & 0 \\
                 0 & 0 & T_{yy} & 0 \\
                 0 & 0 & 0 & T_{zz}
\end{pmatrix}
= \frac{\pi^2}{720\,a^4}\;
\begin{pmatrix} -1 & 0 & 0 & 0 \\
                 0 & 1 & 0 & 0 \\
                 0 & 0 & 1 & 0 \\
                 0 & 0 & 0 & -3
\end{pmatrix}
\label{eq:cas_stress}
\end{equation}
where the plates are separated by a distance $a$ in the $z$-direction. 
Thus there is a constant
negative energy density between the plates. 

\begin{figure}
\psfig{file=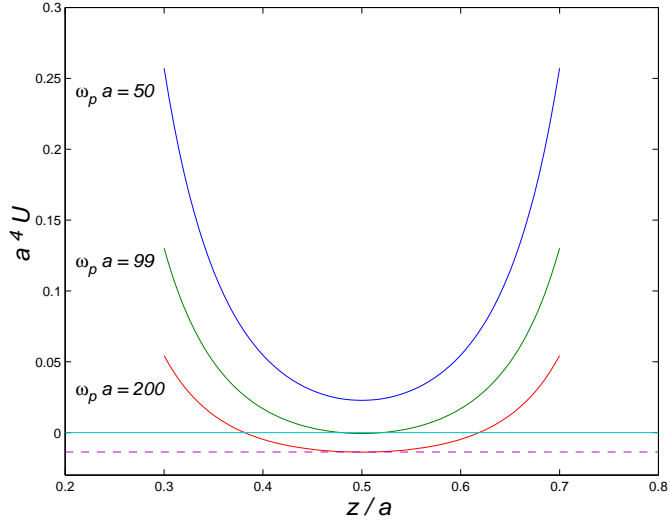,width=3.5in}
\caption{The energy density in the vacuum region between two 
dielectric   
half-spaces is illustrated for three values of the   
parameter $\protect\omega_{p}a.$ The dashed horizontal line 
is the energy 
density for the perfectly conducting limit. }
\label{fig:cas}
\end{figure}

The case of plates of
finite reflectivity is more complicated, and the local stress tensor
can no longer be recovered simply from knowledge of the force per unit
area. The reason for this is that symmetry under Lorentz boosts
parallel to the plates, a key ingredient in the argument leading to
Eq.~(\ref{eq:cas_stress}), no longer holds. Helfer and Lang~\cite{HL99}
have noted that now there could be a positive self-energy density
associated with each plate, even when the plates are widely separated.
In this case, an attractive force is no guarantee of negative energy
density. Consider for example the case of classical electrostatics, where
opposite charges attract, but the local energy density, proportional
to the squared electric field, in non-negative. In the case of
plates described by a plasma model dielectric function, the local
energy density between the plates was calculated in Ref.~\refcite{SF02},
with the results illustrated in Fig.~\ref{fig:cas}.

The basic result is that the local energy density at the center
of the region between the plates will become negative when the 
reflectivity of the plates is sufficiently large, specifically
when $\omega_{p} a > 100$, where $\omega_{p}$ is the plasma frequency.
Thus local negative energy density is possible in the Casimir
effect, but is not inevitable.

\section{Negative Energy from Quantum Coherence Effects}

Another way to create local negative energy densities is with
non-classical states of the quantized electromagnetic field,
or indeed any quantum field. It was proven many years ago by
Epstein {\it et al}~~\cite{EGJ} that all quantum field theories
contain quantum states in which the local energy density at a given
point may be made negative. In fact, it can be arbitrarily 
negative~\cite{Fewster05}. Explicit examples of this
phenomenon are rather easy to construct, and include the moving
mirror models of Fulling and Davies~\cite{FD76} and the squeezed
vacuum states~\cite{Caves}. The energy density in several
non-classical states, including squeezed vacua, is discussed, for
example, in Refs.~\refcite{BFR02,FR08}. 

 In the case of a single monochromatic mode,
the energy density will be oscillatory, with both negative and
positive energy density intervals, but with the time-averaged energy 
being positive. By making the frequency of the mode arbitrarily high,
one can make the peak values of both the negative and positive energy
densities arbitrarily large, but the duration of
the period of negative energy will become arbitrarily short. 
As we will see in 
Sect.~\ref{sec:QI}, this is a very general feature of negative
energy from quantum coherence effects. 

Unrestricted negative energy would have dramatic and disturbing
consequences, including violations of the second law of 
thermodynamics~\cite{F78}, formation of naked singularities in
general relativity~\cite{FR90,FR92}, traversable 
wormholes~\cite{MTU,Visser}, and faster than light 
travel~\cite{Alcubierre}. The later two possibilities could allow the
creation of a time machine, with all of the logical problems inherent
in time travel.

\section{Quantum Inequalities}
\label{sec:QI}

 Unrestricted negative energy would cause serious
problems for physics as it is presently formulated. This leads to
the suspicion that the laws of physics do place restrictions on
negative energy and related effects. This is indeed the case. In
the case of free fields in Minkowski spacetime, it is possible to
prove ``quantum inequalities'' which greatly constrain the magnitude
and duration of negative energy fluxes or periods of negative energy
density. The first versions~\cite{F78,F91} of the quantum inequalities 
were limits on the negative energy fluxes, which showed that the
duration of such a flux has an inverse relation to its magnitude. 
This limit is sufficient to show that macroscopic violations of the
second law are not possible. Later 
authors~\cite{FR95,FR97,FE,FLAN,Fewster}
proved quantum inequalities for the expectation value of the energy
density in arbitrary quantum states. Let
\begin{equation}
 \rho(t) = \langle T_{tt}(t) \rangle
\end{equation}
be the expectation value of the energy density operator in an arbitrary
state, evaluated on the worldline of an inertial observer at time $t$.
Further, let $g(t,\tau)$ be a sampling function in $t$ with a
characteristic width of $\tau$. The quantum inequalities for a massless
field  take the general form
\begin{equation}
\int \rho(t)\, g(t,\tau)\, dt \geq -\frac{C}{\tau^d} \,,     \label{eq:QI}
\end{equation}
where $d$ is the spacetime dimension, and $C$ is a positive constant.
The basic physical content of Eq.~(\ref{eq:QI}) is that a observer
who sees negative energy lasting for a time of order $\tau$ will 
measure the magnitude of this negative energy density to be bounded
by about $C/\tau^d$. 

In the case of a massless scalar field in two-dimensional spacetime, 
$d=2$, Flanagan~\cite{FLAN}
has obtained the optimum bound to be given by
\begin{equation}
C = \frac{\tau^2}{24 \pi} \int_{-\infty}^\infty dt
\frac{{\dot{g}}^2}{g}\,,
\label{eq:QI2D}
\end{equation} 
where $\dot{g} =\partial g/\partial t$. For example, for the case of
a Gaussian sampling function,
$g(t,\tau) = (\sqrt{\pi}\,\tau)^{-1} \, {\rm e}^{-t^2/\tau^2}$,
one finds
$C = {1}/({12 \pi})$ .
This bound will play a key role in the results on energy density
fluctuations to be discussed in Sect.~\ref{sec:STfluct}.
Flanagan's bound, Eq.~(\ref{eq:QI2D}), is optimal in the sense that
one can construct a quantum state for which this relation is an
equality, thus proving that there cannot be a more restrictive bound
for arbitrary states. In four dimensional spacetime, the optimal
bound is not known, but Fewster and Eveson~\cite{FE} have proven
an inequality for a general sampling function. 
In this case, typical values of $C$ are of order $10^{-3}$ for
functions such Lorentzians or Gaussians.

It may be shown\cite{F78,FR90,FR92,FR97,PF97,TRMGM} that these 
inequalities, and their curved spacetime
analogs, greatly restrict the effects of negative energy, making
macroscopic violations of the second law or of cosmic censorship
unlikely, and preventing the construction of macroscopic wormholes 
or warpdrives.

\section{Possible Experiments to Detect Negative Energy?}
\label{sec:exp}

Although negative energy and related sub-vacuum effects are constrained
by quantum inequalities, this does not mean that they are unobservable
in principle. Unfortunately, the gravitational effects of negative
energy are extremely small. However, there is still a possibility of a
laboratory experiment using non-gravitational effects. The role of
quantum inequalities in quantum optics has been discussed in
Refs.~\refcite{M,MS}.

An early
attempt to devise an experiment was made in Ref.~\refcite{FGO92}.
These authors considered a spin system in a magnetic field interacting
with the quantized electromagnetic field in a non-classical state, such
as a squeezed vacuum, for a single plane wave mode. 
Normally, photons will flip spins and cause a decrease
 in the magnetization of the
system. However, in a non-classical photon state, the instantaneous
magnetization can actually increase above its value in the
vacuum. In the model of  Ref.~\refcite{FGO92}, this increase
occurs during the interval when the expectation value of energy 
density of the quantized electromagnetic field is negative. One
can interpret this result as follows: quantum vacuum fluctuations
cause some de-alignment of the spins, compared to what would occur
in a world without vacuum fluctuations. One cannot turn off these
fluctuations, but they can be momentarily suppressed, resulting in
``repolarization''. 
This effect seems too small for a realistic experiment. It
should also be noted that it is really measuring the
mean squared magnetic field $\langle B^2 \rangle$ in the non-classical
state, rather than the energy density. It is only in the special case
of a plane wave mode, where  $\langle B^2 \rangle =\langle E^2
\rangle$, that this effect is a measure of negative energy density.

Another proposal to measure sub-vacuum effects was made in
Ref.~\refcite{HF08}, where the effect of squeezed states of the
photon field on electron interference was discussed. Normally,
the scattering of photons by the electrons will lead to dephasing, 
decreasing the contrast of the 
interference pattern. However, when the photons are in a squeezed
vacuum state, and the interference pattern is formed only from 
electrons which pass through the interferometer at selected times
in the cycle of the excited mode, then the contrast can increase
compared to the vacuum case. This effect of ``recoherence'' is
similar to the  repolarization effect for spins.
 Vacuum fluctuations of the
electromagnetic field cause fluctuations of the Aharonov-Bohm phase
of the electrons, which in turn lead to a loss of contrast,
compared to what would be seen in a world without vacuum
fluctuations. Again, the best that we can do is to momentarily
suppress these fluctuations, and cause a small increase in contrast.
This effect can be expressed as a change in a double surface integral
of $\langle F_{\mu\nu}(x)\,F_{\alpha\beta}(x') \rangle$, the electromagnetic
field strength correlation function, with the integration taken over
a surface bounded by the electron paths. This is not a local quantity
like the energy density, but suppression of the Aharonov-Bohm phase
fluctuation is a sub-vacuum effect, just as is negative energy density.
The effect of recoherence is in principle observable, but probably
not with current technology.
 
Another potentially observable effect of sub-vacuum phenomena arises
from the spontaneous decay rates of atoms. It is well known that
vacuum fluctuations are essential for  spontaneous decay, because
atomic energy levels would be eigenstates of the Hamiltonian, and
hence stable, were it not for the coupling to the quantized
electromagnetic field. It was recently proposed in Ref.~\refcite{FR09}
that non-classical states of the photon field might lead to observable
suppression of atomic decay rates. The proposed experiment involves
sending a beam of atoms in an excited state through a cavity, in which
one mode is excited in a non-classical state, such as a squeezed
vacuum. On average, the effect of the excited state will be to
increase the decay rate, as would be expected from stimulated emission
effects. However, if the atoms pass through the cavity at certain
times in the cycle of the excited mode, then the decay probability
during the transit period can be reduced compared to the case when the
electromagnetic field in the cavity is in the vacuum state. As in the
effects discussed in the previous two paragraphs, this can be
interpreted as suppression of the usual vacuum fluctuation effects.
Under certain conditions, this effect can serve to measure
$\langle E^2 \rangle$, the shift in mean squared electric field due
to the non-classical state (but not including Casimir effects due
solely to the cavity). In the case where the frequency of the 
cavity field is near the atomic transition frequency, the transit time
is small compared to the associated period, and the mode function is 
approximately constant along the atom's path, then the decay probability
is
\begin{equation}
\frac{P}{P (0)} =1+ \frac{1}{ {f^2} ({\bf x_0})} \, 
 \langle  E^2 ({\bf x_0}, t) \rangle  \,.
\label{eq:P2/P20_1}
\end{equation}
Here $f^2 ({\bf x_0})$ is the squared mode function for the
excited mode, evaluated at a point ${\bf x_0}$ on the atom's path,
 $P$ is the decay probability for the non-classical
state, and $P(0)$ is that for the vacuum state. When 
$\langle  E^2 ({\bf x_0}, t) \rangle <0$, then $P < P(0)$, and
the decay rate has been suppressed. In Ref.~\refcite{FR09}, a quantum
inequality is proven which states that
\begin{equation}
\langle  E^2 ({\bf x_0}, t) \rangle \geq - f^2 ({\bf x_0}) \,,
\end{equation}
thus guaranteeing that $P \geq 0$, as required. However, a state which
comes close to saturating the quantum inequality bound will lead to a
significant fractional decrease in decay probability when the atom is
in the cavity. Note that this is quite different from the more
familiar suppression of atomic decay rates in cavities compared to
empty space, which can be interpreted in terms of a lack of available
modes in to which to atom can decay. The decay probability during
transit is small to begin with, of order $10^{-8}$ in some examples
treated in Ref.~\refcite{FR09}, so a large number of atom needs to be
used to produce a statistically significant result. However, a
realistic experiment might be feasible, and is currently under study.

\section{Quantum Stress Tensor Fluctuations}
\label{sec:STfluct}

So far, we have been discussing situations where the expectation value
of the energy density can be negative. However, there is another
sense in which negative energy density can arise in quantum field 
theory. This is when quantum fluctuations momentarily create a region
of negative energy. This can occur even when the expectation value is 
non-negative. A simple example is the Minkowski vacuum state, where
$\langle T_{\mu\nu} \rangle =0$, but the state is not an eigenstate of 
$T_{\mu\nu}$. This means that there must be both positive and negative
fluctuations and an associated probability distribution function. 
To find this, we need the probability distribution for a stress
tensor operator averaged over  a sampling function
in time, or spacetime.

\begin{figure}
\psfig{file=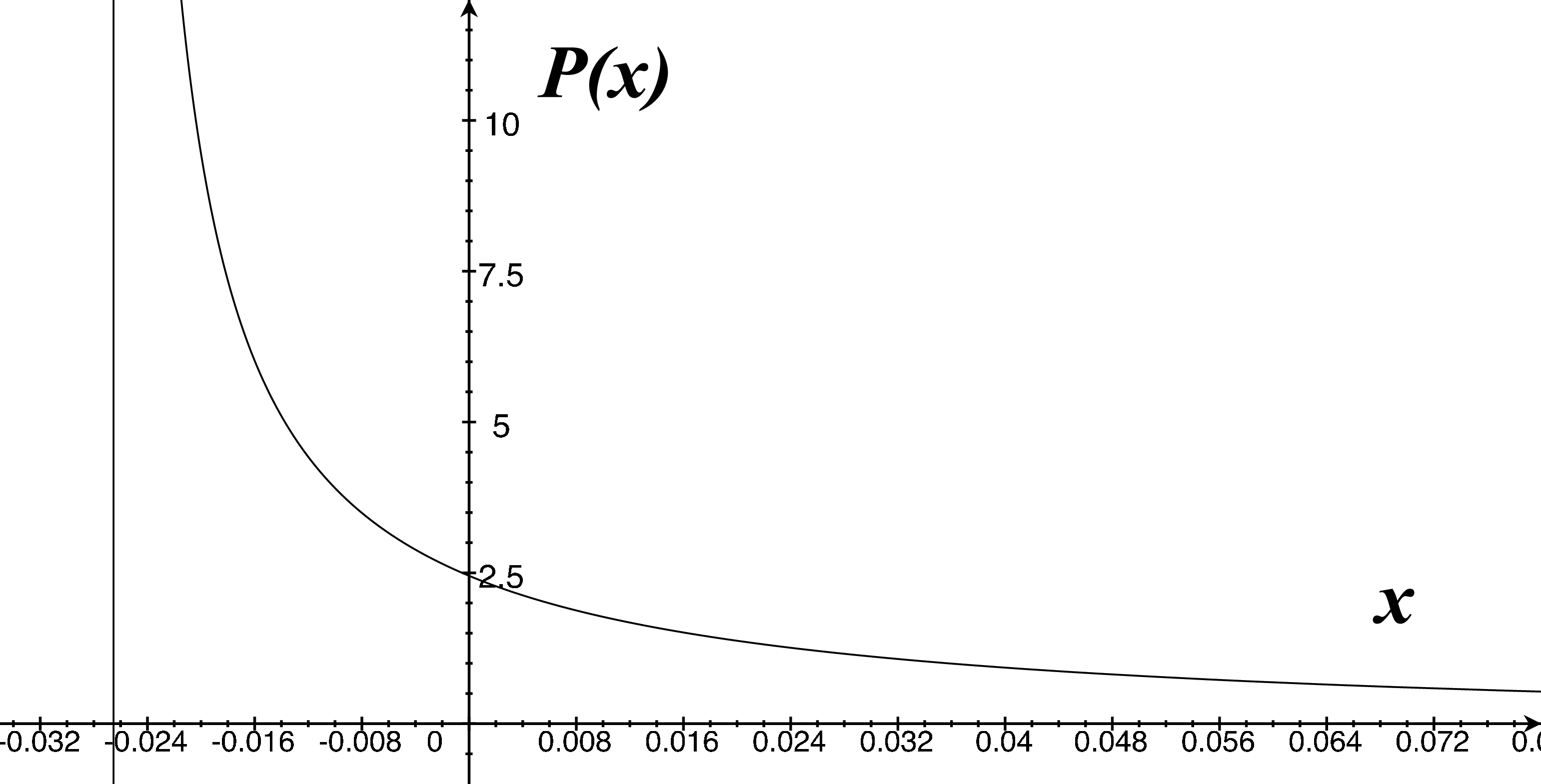,width=4.0in}
\caption{The probability distribution $P(x)$ for the smeared energy density
  of a massless scalar field in two-dimensional Minkowski spacetime is
plotted. Here $x = u\, \tau^2$, where $u$ is the energy density
operator averaged in time with a Gaussian function of width
$\tau$. The lower limit of $P(x)$ occurs at $x=-x_0= -1/(12 \pi)$,
illustrated by the vertical line. }
\label{fig:prob}
\end{figure}

In general, this is still an unsolved problem. However, it has
recently been solved for the case of two-dimensional conformal field 
theory with a Gaussian average 
in time~\cite{FFR09}.  Here we simply quote the results, which will be
derived in Ref.~\refcite{FFR09}, and discuss some of their physical
implications. Let $T_{tt}$ be the energy density operator, and define
the averaged energy density by
\begin{equation}
u =\frac1{\sqrt{\pi}\,\tau} 
\int_{-\infty}^{\infty} T_{tt}(t,x)\, {\rm e}^{-t^2/\tau^2} \, dt  \,.
\end{equation}
The associated probability distribution is a Gamma distribution given
by 
\begin{equation}
P(x) = \frac{\pi^{c/12}}{\Gamma(c/12)}\, (x +x_0)^{c/12 -1} \;
{\rm e}^{-\pi(x+x_0)}\, , \qquad x>-x_0 \,,
\end{equation}
and $P(x) =0$ for $x<-x_0$. Here $x = u\, \tau^2$, $c$ is the central
charge, and $-x_0/\tau^2$ is the quantum inequality bound on
expectation values of $T_{tt}$ in arbitrary quantum states. The
probability distribution is plotted in Fig.~\ref{fig:prob} for the
case of a free massless scalar field in two-dimensional Minkowski
spacetime, for which $c=1$. 

The lower cutoff at the quantum inequality bound is expected to be a
general feature of the probability distribution for vacuum
fluctuations of an averaged energy density. It is of interest to note 
that $0.84$ of the
area of the graph in Fig.~\ref{fig:prob} lies to the left of the
origin. This means that
a measurement of the averaged energy density will find a negative
result $84\%$ of the time. However, when a positive value is found, 
it is typically larger in magnitude.

The probability distribution for the energy density in 
four-dimensional theories has not yet been found, but is of considerable
interest. One application is to inflationary cosmology, where
quantum stress tensor fluctuations might contribute a potentially
observable component to the cosmological density
fluctuations~\cite{WKF07}. This component would be non-Gaussian
in a way which is associated with the skewness of the quantum stress
tensor probability distribution. A distinct, more exotic,
application to cosmology arises in models which employ anthropic
reasoning to compute probabilities of various observables.
These models require a counting of observers, usually assumed to
be beings like ourselves in the sense of having arisen from biological
evolution on earth-like planets. However, another possibility are
``Boltzmann brains'' which have nucleated from the vacuum in deSitter
or even Minkowski spacetime.  Quantum energy fluctuations can
sometimes produce large concentrations of energy, which very
occasionally might be capable of conscious  thought. The probability
per unit volume of this is very tiny, but the available volume
is vastly larger than that for biological systems.
If ``Boltzmann brains'' are  the more
prevalent type of observer, it would greatly complicate attempts at anthropic
prediction. (For further discussion and references,
see, for example Refs.~\refcite{GV08,DGLNSV}.)
The key to studying this question is in the details of the
long positive tail of the probability distribution, which can tell
us how likely the appearance of a ``Boltzmann brain'' in a given
region might be. 

\section{Summary}

We have seen that quantum field theory allows local negative energy
and other sub-vacuum effects. These effects are strongly restricted by
quantum inequalities, but are nonetheless potentially observable.
We have reviewed some proposals to measure these effects in laboratory
experiments, the most promising of which involves changes in the decay
rates of atoms, and could conceivably lead to measurements of negative
mean squared electric fields. We have also discussed some new results
on the probability distribution for vacuum stress tensor fluctuations. This
distribution has a lower cutoff at the quantum inequality bound on the
expectation value in an arbitrary state, but has a tail in the positive
direction. Thus a typical fluctuation in the local energy density is
negative but bounded below, while rare but extremely large positive 
fluctuations are possible. Stress tensor fluctuations effects could
be important in inflationary cosmology.

\section*{Acknowledgments}
I would like to thank Chris Fewster and Tom Roman for valuable
discussions.
This research was supported in part by the US National Science
Foundation under Grant No. PHY-0855360.

\end{document}